\documentstyle[a4,12pt,epsfig]{article}
\begin{document}
\large
\begin{center}
{\bf GRB 000926 and its optical afterglow: Another evidence for non-isotropic
 emission }
\end {center}

\normalsize

\begin{center}
{\bf R. Sagar$^{1}$, S. B. Pandey$^{1}$, V. Mohan$^{1}$, D. Bhattacharya$^{2}$ 
   and A. J. Castro-Tirado$^{3}$} \\

\medskip
 {\it $^{1}$U. P. State Observatory, Manora Peak, Nainital -- 263 129, India}\\
 {\it $^{2}$Raman Research Institute, Bangalore -- 560 080, India}\\
 {\it $^{3}$IAA-CSIC, P.O. Box 03004, E-18080, Granada, Spain}
\end {center}

\bigskip

\begin{abstract}

The CCD Johnson $BV$ and Cousins $RI$ photometric magnitudes are determined for 
20 stars in the field of GRB 000926. They are used to calibrate the present 
$R$ as well as published $BVRI$ photometric magnitudes of the GRB 000926 
afterglow. Optical light curves of the afterglow 
emission are obtained in $B$ and $R$ passbands. They show a steepening of the 
flux decay as expected of an anisotropic fireball losing collimation with the 
fall of the bulk Lorentz factor. We derive the early and late time flux decay 
constants as $1.4\pm0.1$ and $2.6\pm0.06$ respectively. Steepening in the flux 
decay seems to have started around 1.7 days after the burst. Negligible Galactic
but relatively large intrinsic extinction amounting $E(B-V) =$ 0.03 and 
0.36$\pm$0.02 mag respectively are derived in the direction of GRB 000926. The 
value of the spectral index in the \mbox{X-ray}--optical--near-infrared region 
is $\sim -0.9$. The determination of the redshift $z=2.0369$ indicates a 
cosmological origin of the burst at a luminosity distance of 16.6 Gpc. The 
observed fluence in the energy range 20--100 keV indicates, if isotropic, the 
release of $\ge 10^{53}$ ergs of energy. Attributing the observed break in the 
light curve at 1.7 days to the onset of sideways expansion of a jet-like ejecta,
we infer an initial jet opening angle of $\sim$0.14~radian. This indicates 
a large anisotropy in the original emission and the amount of released energy 
is reduced by a factor of $\sim$ 100 relative to the isotropic value, 
which can be understood in terms of the currently popular stellar death models. 
\end{abstract}


\section {Introduction}

Multiwavelength follow-up of Gamma-Ray Burst (GRB) afterglows has
revolutionized GRB astronomy in recent years, yielding a wealth of
information about the nature and origin of GRBs (cf. Klose 2000; 
Kulkarni et al. 2000; Galama 2000; Castro-Tirado et al. 1999 and 
references therein).  One of the most important clues to the origin
of GRBs is the total amount of energy released in the event.  Optical
observations have helped to derive this important quantity in a number
of cases by setting the distance scale through measurement of redshift,
and by determining the degree of collimation of the initial emission
through detailed study of the afterglow light curve.  Clear evidence
of the steepening of the light curve, signature of an originally collimated
outflow expanding laterally, has been found within a few days of the GRB
event in several afterglows (Castro-Tirado et al 1999; Kulkarni et al 1999;
Stanek et al. 1999; Halpern et al. 2000c; Sagar et al. 2000b)

As a part of an international collaborative programme coordinated
by one of us (AJCT), the Uttar Pradesh State Observatory (UPSO), Naini Tal
has been involved in a regular programme of GRB follow-up since 
January 1999.  So far successful photometric observations of five afterglows 
have been carried out at UPSO.  The results of four of these have been
reported earlier (Sagar et al. 1999, 2000a, 2000b).  In this paper
we report the studies made on the fifth, namely, GRB~000926.

GRB~000926 was detected on 2000 September 26 at 23:49:33 UT by the 
Inter-Planetary Network (IPN) group of spacecrafts
Ulysses, Russian Gamma-Ray Burst Experiment (KONUS) and Near Earth Asteroid 
Rendezvous (NEAR) (Hurley et al. 2000). The burst had a \mbox{25--100~keV} 
fluence of $\sim 2.2 \times 10^{-5}$ erg cm$^{-2}$. The total duration of the 
GRB is $\sim 25$~s, putting it in the class of ``long-duration'' bursts, to 
which all GRBs with detected afterglows so far belong (cf. Lamb 2000). 

Optical observations taken on 2000 September 27 by Gorosabel et al. (2000) 
and Dall et al. (2000) detected independently a new bright point like source 
at $\alpha_{2000}=17^h 04^m 09.^s7; \delta_{2000}=+51^{\circ} 47^{'} 10^{''}$ 
with a strong candidacy for the afterglow from GRB 000926. At the same 
location, Frail \& Berger (2000) detected a new radio source with VLA at 
8.46 GHz and 4.86 GHz. Coincident (within errors) with the location of optical 
and radio afterglows, Piro \& Antonelli (2000) and Garmire et al. (2000) 
detected the \mbox{X-ray} afterglow of GRB 000926. The near-IR $J$ detection of 
the afterglow has been reported by Di Paola et al. (2000), while observations 
in $K^{'}$ passband by Kobayashi et al. (2000) clearly detect the 
Optical Transient (OT). 

The high resolution optical spectrum of the OT of GRB 000926 taken on 2000 
September 29.26 UT (cf. Castro et al. 2000) indicate a redshift value of
$z = 2.0369\pm0.0007$. This is a refinement of $z = 2.066$ reported earlier 
by Fynbo et al. (2000c) using the low resolution optical spectrum 
taken on 2000 September 27.91 UT. Fynbo et al. (2000d, e) report
the detection of the host galaxy with an R magnitude of $\sim 24$.

In order to provide  accurate photometric magnitudes of the OT, we also imaged 
the field of GRB 000926 along with SA 110 region of Landolt (1992). A total of 
20 stars in the field have been calibrated and their standard $BVRI$ magnitudes 
are given here. We present the details of our optical observations in the 
next section, and discuss the light curves and other results in the remaining 
sections.

Recently Price et al (2000b) have presented an independent series of 
observations in multiple optical/IR bands. The results obtained by them 
are consistent with those presented here.

\section { Optical observations, data reduction and calibrations } 

The optical observations of the GRB 000926 afterglow were carried out on 
2000 September 29. We used a 2048 $\times$ 2048 pixel$^{2}$ CCD system 
attached at the f/13 Cassegrain focus of the 104-cm Sampurnanand telescope of 
UPSO, Nainital. One pixel of the CCD chip corresponds to 0.$^{''}$38, and the 
entire chip covers a field of $\sim 13^{'} \times 13^{'}$ on the sky. 
The CCD $BVRI$ observations of the GRB 000926 field along with 
Landolt (1992) SA 110 region have also been carried out on 27th October 2000 
for calibration purposes during photometric sky conditions. The log of CCD 
observations is given in Table 1. In addition to these observations, several 
twilight flat field and bias frames were also observed.

\begin{table}
{\bf Table 1.}~Log of CCD observations of GRB 000926 and SA 110 
Landoldt (1992) standard fields.

\begin{center}
\begin{tabular}{clcl} \hline 
 Date & Field  & Filter & Exposure (in seconds) \\ \hline
 29/30 Sep 2000 & GRB 000926 & $R$ & 600$\times$3 \\
 27/28 Oct 2000 & GRB 000926 & $B$ & 300$\times$3 \\
 27/28 Oct 2000 & GRB 000926 & $V$ & 200$\times$3 \\
 27/28 Oct 2000 & GRB 000926 & $R$ & 150$\times$3 \\
 27/28 Oct 2000 & GRB 000926 & $I$ & 150$\times$3 \\
 27/28 Oct 2000 & SA 110 & $B$ & 150$\times2, 100\times$4 \\
 27/28 Oct 2000 & SA 110 & $V$ &  100$\times2, 50\times$4 \\
 27/28 Oct 2000 & SA 110 & $R$ &30$\times$5 \\ 
 27/28 Oct 2000 & SA 110 & $I$ & 30$\times$5 \\  \hline
\end{tabular}
\end{center}
\end{table}

The CCD frames were cleaned using standard procedures. Image processing was done
using ESO MIDAS and DAOPHOT softwares. Atmospheric extinction coefficients were 
determined from the observations of the brightest star present in the SA 110 
field and these are used in further analysis. Standard magnitudes of 8 stars in 
the field of SA 110 were taken from Landoldt (1992). They cover a wide range in 
colour ($0.5 < (V-I) < 2.6$) as well as in brightness ($11.3 < V < 14.2$). 
The transformation coefficients were determined by fitting least square linear 
regressions to the standard $BVRI$ photometric indices as function of the 
observed instrumental magnitudes normalised for 1 second exposure time.  The
following colour equations were obtained for the system.

\noindent $(B-V) = (1.17\pm0.02) (b-v)_{CCD} - (0.14\pm0.02) $  \\
	$(V-R) = (0.91\pm0.02) (v-r)_{CCD} - (0.39\pm0.02) $  \\
	$(V-I) = (0.92\pm0.02) (v-i)_{CCD} + (0.01\pm0.03) $  \\
	$ V   = v_{CCD} - (0.02\pm0.01) (B-V) - (4.62\pm0.01) $  \\
where $v_{CCD}, (b-v)_{CCD}, (v-r)_{CCD} $ and $(v-i)_{CCD}$
represent the instrumental colour indices corrected for atmospheric extinction.
The errors in the colour coefficients and zero points are obtained
from the deviation of data points from the linear relation.

For increasing the photometric precision of fainter stars, the data are 
binned in $2 \times 2$ pixel$^2$ and also all the CCD images of GRB 000926 
field taken on a day are co-added in the same filter. From these added 
images, profile-fitting magnitudes are determined using DAOPHOT software. The 
standard magnitudes of the stars are determined using the above transformations.
The magnitude of the OT of GRB 000926 determined in this way is $R = 21.7\pm0.2$
at 2000 September 29.617 UT. $BVRI$ photometric magnitudes of 20 stars in the
GRB 000926 field are listed in Table 2. The present $B$ and $R$ magnitudes of 
star 19 (comparison star) agree within errors with an independent determination
by Halpern et al. (2000a) indicating that present photometric calibration is
secure. Fig. 1 shows the location of the GRB 000926 afterglow and the 
photometered stars on the CCD image taken from UPSO, Nainital. 

The above photometrically calibrated magnitudes have also been used for 
calibrating other photometric measurements 
of GRB 000926 afterglow published in the GCN circular by Fynbo et al. (2000a,
b, d), Halpern et al. (2000a, b), Hjorth et al. (2000a, b), 
Price et al. (2000a),
Rol et al. (2000), Veillet (2000) and Vrba \& Canzian (2000) by the time of 
paper submission. In order to avoid errors arising due to different photometric 
calibrations, we have used only those published $BVRI$ photometric measurements 
whose magnitudes could be determined relative to the stars given in Table 2.
For converting the Sloan $i$ measurements given by Rol et al. (2000) into 
Cousins $I$ system we also used the calibrations given by Fukugita et al. 
(1995). The $J$ magnitude is adopted from Di Paola et al. (2000).
A total of 53 photometric data points in five passbands with the distribution 
of points: $N(B,V,R,I,J) = (7, 4, 37, 4, 1)$ are there for our analysis.

\begin{figure}
\begin{center}\vspace*{-3.8cm}
\hspace*{-1cm}
\epsfig{file=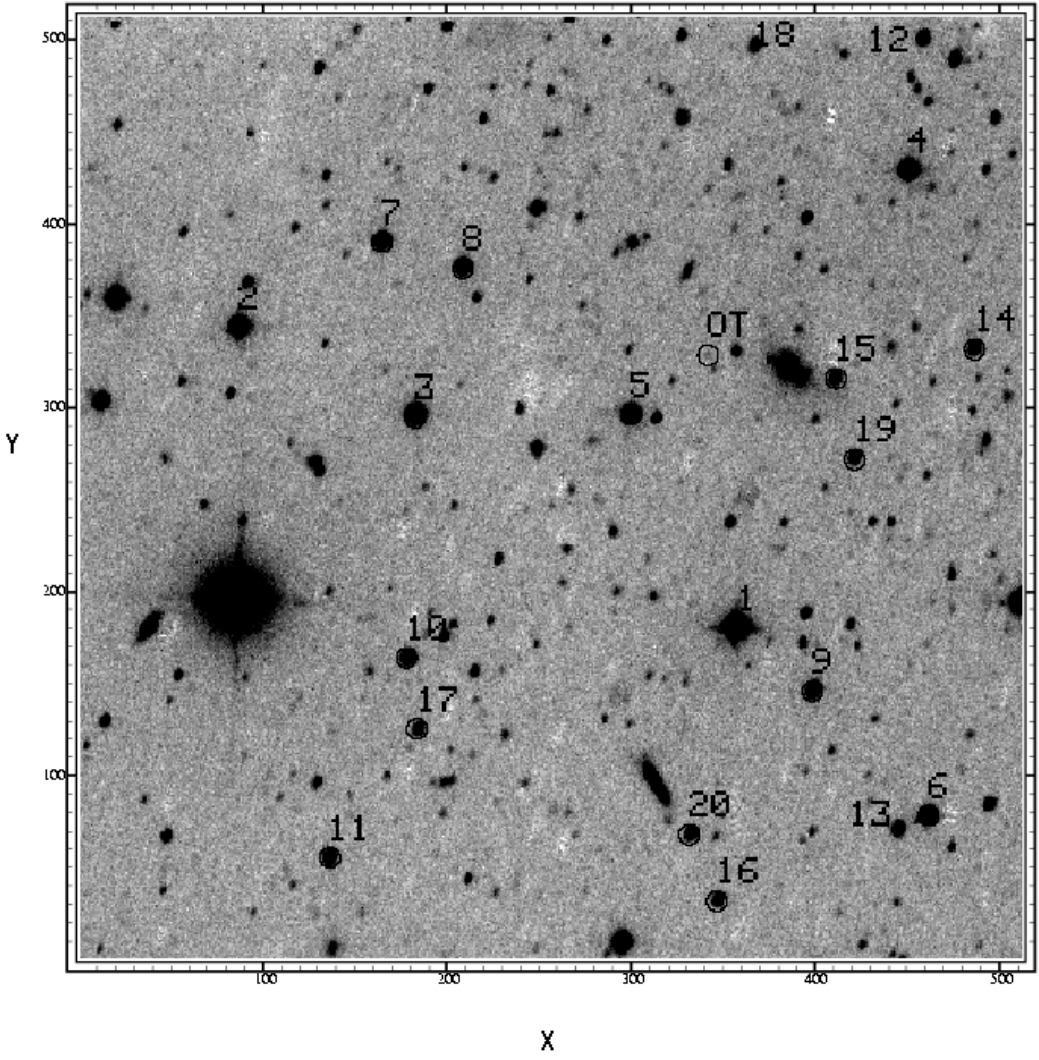,width=\textwidth}
\end{center}
\vspace*{-1.0cm}
\label{ident} 
\caption{Finding chart for GRB 000926 field is produced from the CCD images 
taken from UPSO, Nainital on 2000 September 29.6 UT in $R$ filter with exposure 
time of 30 minutes. North is up and East is left.  The optical transient (OT) 
and the stars with present $BVRI$ magnitudes are marked. The ($X, Y$) are the 
pixel coordinates on a scale of $0.^{''}76$ per unit. The corresponding sky 
coordinates are $\Delta\alpha$ = 4255$\pm$0.4 - (1.554$\pm$0.001)$X$ +  
(0.068$\pm$0.001)$Y$ and $\Delta\delta$ = 340$\pm$0.6 + (0.041$\pm$0.001)$X$ + 
(0.962$\pm$0.001)$Y$, where $\Delta\alpha$ and $\Delta\delta$ are offsets in 
arcsec with respect to RA=17h and Dec=51.5~deg respectively.} 
\end{figure}

\footnotesize
\begin{table} 
{\bf Table 2.}~The identification number(ID), Right Ascension ($\alpha$)
and Declination ($\delta$) for epoch 2000 of the stars in the region of GRB 
000926 are taken from the catalogue of Monet (1997). Standard $V, (B-V), (V-R)$ 
and $(R-I)$ photometric magnitudes and their associated DAOPHOT errors 
($\sigma$)  of the stars are also given.  These errors are primarily related 
to signal to noise ratio and do not include the errors related to colour 
transformations and magnitude zero points. Star 19 is the comparison star
mentioned by Halpern et al. (2000a).

\begin{center}
\begin{tabular}{ccc cc cc cc cc} \hline 

 ID & $\alpha_{2000}$ & $\delta_{2000}$ & $V$& $\sigma_V$  
& $B-V$ & $\sigma_{B-V}$ & $V-R$ & $\sigma_{V-R}$ & $V-I$ &$\sigma_{V-I}$ \\ \hline
1 &$17^{h} 4^{m} 07.^{s}50$&$51^{\circ}44^{'}49.^{''}3$& 13.58& 0.01 &0.83 &0.01
& 0.54& 0.01& 0.88& 0.01\\
2 &17 4 36.19& 51 47 15.0& 14.41& 0.01 &0.56 &0.01& 0.42& 0.01& 0.67& 0.01 \\
3 &17 4 26.06& 51 46 33.5& 14.64& 0.01 &0.63 &0.01& 0.46& 0.01& 0.74& 0.01 \\
4 &17 3 58.86& 51 48 52.1& 14.82& 0.01 &0.62 &0.01& 0.44& 0.01& 0.72& 0.01 \\
5 &17 4 13.96& 51 46 35.9& 15.31& 0.01 &1.08 &0.01& 0.71& 0.01& 1.23& 0.01 \\
6 &17 3 56.25& 51 43 13.7& 15.50& 0.01 &0.73 &0.02& 0.45& 0.02& 0.75& 0.02 \\ 
7 &17 4 28.41& 51 48 02.4& 15.55& 0.01 &0.65 &0.01& 0.45& 0.01& 0.75& 0.01 \\
8 &17 4 23.81& 51 47 50.9& 15.70& 0.01 &0.45 &0.01& 0.35& 0.01& 0.60& 0.01 \\ 
9 &17 4 03.04& 51 44 17.4& 15.79& 0.01 &0.57 &0.01& 0.42& 0.01& 0.71& 0.01 \\
10 &17 4 25.93& 51 44 25.6& 16.29& 0.01 &0.74 &0.01& 0.41& 0.02& 0.75& 0.02 \\
11 &17 4 29.79& 51 42 39.2& 16.72& 0.01 &0.87 &0.02& 0.50& 0.02& 0.87& 0.02 \\
12 &17 3 58.40& 51 50 00.6& 16.83& 0.01 &0.68 &0.02& 0.42& 0.02& 0.79& 0.02 \\
13 &17 3 57.87& 51 43 06.6& 16.93& 0.02 &0.68 &0.02& 0.36& 0.02& 0.65& 0.02 \\
14 &17 3 54.77& 51 47 20.2& 17.16& 0.01 &0.67 &0.02& 0.39& 0.01& 0.70& 0.01 \\
15 &17 4 02.50& 51 47 01.7& 17.54& 0.01 &1.09 &0.02& 0.66& 0.02& 1.15& 0.02 \\
16 &17 4 07.82& 51 42 25.7& 17.58& 0.01 &0.56 &0.02& 0.32& 0.02& 0.59& 0.02 \\
17 &17 4 25.11& 51 43 48.4& 17.62& 0.01 &0.85 &0.02& 0.47& 0.02& 0.85& 0.02 \\
18 &17 4 07.93& 51 49 53.2& 17.62& 0.01 &0.77 &0.02& 0.45& 0.02& 0.79& 0.02 \\
19 &17 4 01.21& 51 46 19.9& 17.63& 0.01 &0.86 &0.02& 0.54& 0.01& 0.94& 0.02 \\
20 &17 4 09.58& 51 42 59.9& 17.73& 0.01 &1.08 &0.02& 0.62& 0.01& 1.11& 0.02 \\ \hline
\end{tabular}
\end{center}
\end{table}

\normalsize
\section{ Optical photometric light curves}

We have used the data published in GCN circulars in combination with the 
present measurements to 
study the optical flux decay of GRB 000926 afterglow. Fig. 2 shows a plot of 
the photometric measurements as a function of time. The X-axis is log ($t-t_0$) 
where $t$ is the time of observation and $t_0$ (= 2000 September 26.9927 UT) is 
the time of GRB trigger. All times are measured in unit of day.

\begin{figure*}
\begin{center}\vspace*{-1.5cm}\hspace*{-1cm}
\epsfig{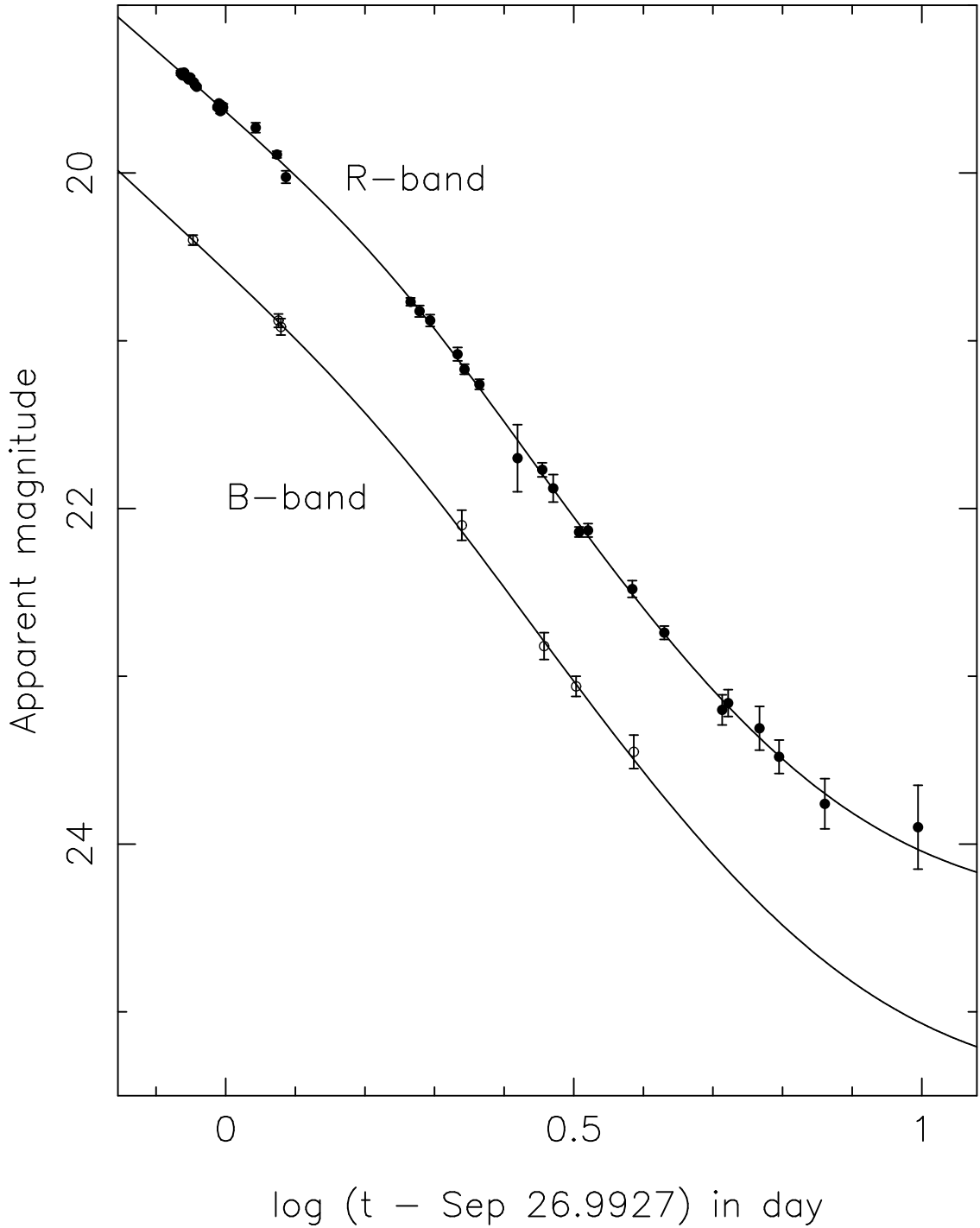}
\end{center}\vspace*{-0.5cm}
\label{light} 
\caption{Light curve of GRB 0000926 afterglow in optical $B$ and 
$R$ photometric passbands.  Flux decay can not be fitted by a single power-law. 
Solid lines represent the least square non-linear fits to the observed data for 
a jet model (see text).} 
\end{figure*}

The emission from GRB 000926 OT is fading in both $B$ and $R$ passbands. 
The flux decay of most of the earlier GRB afterglows, barring a few, 
is generally well characterized by a single power law 
$F(t) \propto (t-t_0)^{-\alpha}$, where $F(t)$ is the flux of the afterglow at 
time $t$ and $\alpha$ is the decay constant. In contrast, the optical light 
curves of GRB 000926 (Fig. 2) can not be fitted by a single power-law. Overall 
the OT flux decay seems to be described by a broken 
power-law as expected in GRB afterglows having jet-like relativistic ejecta 
(Sari et al. 1999; Rhoads 1999). Broken power-law in these light curves can be 
empirically fitted by functions of the form (see Sagar et al. 2000b for details)

$F(t) = 2F_0/[(t/t_b)^{\alpha_1 s} + (t/t_b)^{\alpha_2 s}]^{1/s} + F_g$, 

\noindent
where $F_g$ is the constant flux from the underlying host galaxy, 
$\alpha_1$ and $\alpha_2$ are asymptotic power-law slopes at early and 
late times with $\alpha_1 < \alpha_2$.  The parameter $s (> 0)$ controls the 
sharpness of the break, a larger $s$ implying a sharper break. With $s = 1$, 
this function becomes the same as that used by Stanek et al. (1999) to fit the 
optical light curve of GRB 990510 afterglow. $F_0$ is the flux of afterglow at 
the cross-over time $t_b$. The function describes a light curve falling as 
$t^{-\alpha_1}$ at $t << t_b$ and $t^{-\alpha_2}$ at $t >> t_b$. In jet 
models, an achromatic break in the light curve is expected when the jet makes 
the transition to sideways expansion after the relativistic Lorentz factor 
drops below the inverse of the opening angle of the initial beam. Slightly 
later, the jet begins a lateral expansion which causes a further steepening of 
the light curve. 

We use the observations in $B$ and $R$ to determine the
parameters of the jet model using the above function. We find that the
minimum value of $\chi^2$ is achieved for $s \ge 5$. This indicates that the
observed break in the light curve is sharp, unlike the smooth break
observed in the optical light curve of GRB 990510 (cf. Stanek et al. 1999;
Harrison et al. 1999) but similar to the sharp break observed in the optical
and near-IR light curves of GRB 000301C (cf. Sagar et al. 2000b).  In order to
avoid a fairly wide range of model parameters for a comparable $\chi^2$ due to
degeneracy between $t_b, m_b$ and $m_g$ (magnitudes corresponding to $F_0$ and
$F_g$ respectively), $\alpha_1$, $\alpha_2$ and $s$, we have used a fixed 
value of $s = 5$ in our further analyses.  The least square fit values of the 
parameters $t_b$, $m_b$, $\alpha_1$, and $\alpha_2$ then work out to be 
1.74$\pm$0.11~d, 21.27$\pm$0.15, 1.45$\pm$0.06 and 2.57$\pm$0.10 respectively 
in $R$ band, with a corresponding $\chi^2$ of 0.93 per degree of freedom
($DOF$).  The fit yields the host galaxy contribution to be
$m_g$=24.41$\pm$0.25 in $R$. This is slightly fainter than the value 
$\sim$ 23.9 estimated by Fynbo et al. (2000e) from deep CCD imaging, but 
brighter than $R$=25.19$\pm$0.17 derived from HST measurements 
(Price et al 2000b). 
The main reason for the discrepancy with the HST estimate is possibly the 
restriction of the HST photometry to within a 1.5 arcsec aperture, while no 
such cutoff has been applied to the photometric data used here.  
The derived values of the above parameters do not change significantly with
larger adopted values of $s$, although a marginal decrease in $\chi^2/DOF$ 
is noticed in doing so.  

In $B$ band the available data is sparse, and a fit with fixed $s=5$ and 
$t_b=1.74$~d yields $\alpha_1$=1.55$\pm$0.11, $\alpha_2$=2.72$\pm$0.48 and
$m_g(B)$=24.7$\pm$1.0.  Note the poor constraint on $\alpha_2$ and B-magnitude
of the host galaxy, because of the scarcity of measurements in the late light
curve.  One can improve the constraints on the decay indices somewhat by 
assuming a colour index for the host galaxy, and thereby fixing its B
magnitude from the R magnitude determined above.  The restricted aperture
HST photometry reported by Price et al (2000b) indicate a $B-R$ colour index
of 1.04$\pm$0.53 for the host galaxy.  Fixing the value of $m_g(B)$ at 25.5
using this mean colour index, the fit yields $\alpha_1$=1.53$\pm$0.11
and $\alpha_2$=2.52$\pm$0.12.  On the other hand, if the host galaxy has a 
flat spectral distribution similar to that for GRB~990123 
(Castro-Tirado et al.\ 1999), then $B-R$ colour will be $\sim 0.5$ 
(which is allowed within the errors of the HST measurements and our light 
curve fits), and $m_g(B)$ will be 24.9.  If we use this value for $m_g(B)$, 
then the best fit values of the decay constants become 
$\alpha_1$ = 1.54$\pm$0.10 and $\alpha_2$ = 2.64$\pm$0.12.  We can therefore 
say with confidence that within the range of host galaxy colours allowed by 
the light curves and the HST measurements, the decay indices in $B$ and $R$ 
bands are consistent with each other, and that the break in the light curve 
displays the achromaticity expected in the case of a lateral expansion of 
collimated ejecta.

The best fit light curves for the $R$ and the $B$ bands, with an assumed 
host galaxy colour index of $B-R$=1.04, are displayed in Fig.~2.  The decay 
parameters derived here are consistent with those obtained
by Price et al (2000b) from their measurements.

\subsection{Spectral index of the GRB 000926 afterglow }

We have constructed the GRB 000926 afterglow spectrum on 2000 September 29.25 
UT and 29.76 UT using the available optical, near-IR, radio and \mbox{X-ray} 
observations. 
These epochs were selected for the long wavelength coverage possible at the time
of observations of \mbox{X-ray} flux by Piro \& Antonelli (2000) and Garmire et 
al. (2000). Optical fluxes at the wavelengths of $BVR$ and $I$ passbands have 
been derived using the slope of the fitted light curve shown in Fig. 2 for the 
epochs under consideration. The fluxes for the desired epoch at 8.46 GHz and $J$
wavelength are taken from Frail \& Berger (2000) and Di Paola et al. (2000) 
respectively assuming that the flux decay is similar to optical one. 
We used the reddening map provided by Schlegel, Finkbeiner \& Davis (1998) for 
estimating Galactic interstellar extinction towards the burst and found a small 
value of $E(B-V) = 0.03$ mag. We used the standard Galactic extinction 
reddening curve given by Mathis (1990) in converting apparent magnitudes into 
fluxes and used the effective wavelengths and normalisations by Fukugita et al.
(1995) for $BVR$ and $I$ and by Bessell \& Brett (1988) for $J$. The fluxes thus
derived are accurate to $\sim$ 10\% in optical and $\sim$ 30\% in $J$. Fig. 3 
shows the spectrum of GRB 000926 afterglow from \mbox{X-ray} to radio region. It
is observed that as the frequency decreases the flux increases from \mbox{X-ray}
to radio wavelengths. For a chosen frequency interval, we describe the spectrum
by a single power law: $F_{\nu}\propto\nu^{\beta}$, where $F_{\nu}$ is the flux 
at frequency $\nu$ and $\beta$ is the spectral index. In the \mbox{X-ray} to 
optical region at $\Delta t$ = 2.26 day, the value of $\beta$ is $-0.85\pm$0.02,
while in the \mbox{X-ray} to $J$ region, the value is $-0.88\pm$0.04. At $\Delta
t$ = 2.77 day the value of $\beta$ is $-0.94\pm$0.04 for \mbox{X-ray} to $J$
region. These values agree very well with the spectral slopes derived for the 
\mbox{X-ray} spectrum of the GRB 000926 afterglow by Piro \& Antonelli (2000) 
and Garmire et al. (2000). During $\Delta t$ = 1 to 4 days, the values of $(B-R)
$ and $(B-I)$ are almost the same indicating no change in the optical spectral 
index before and after the time $t_b$. However, the values of $\beta$ in optical
($-1.54\pm0.15$ and $-1.61\pm0.14$ derived at $\Delta t= 0.9$ and 3.9 day  
respectively) and in optical to $J$ region ($-1.68 \pm0.07$ at $\Delta t= 2.26$
day) are much steeper than those derived for the broadband (X-ray to optical)
spectrum. This may indicate presence of 
intrinsic extinction in the GRB 000926 afterglow. The amount of intrinsic 
extinction is determined in the following way assuming that the OT spectrum
follows a single spectral slope from \mbox{X-ray} to near-IR. The observed flux
at \mbox{X-ray} and $J$ wavelengths are used to determine $\beta$ as the
extinction affects them least. This slope predicts the fluxes of GRB 000926 OT
at $BVRI$ wavelengths. The differences between these predicted and the 
corresponding Galactic extinction corrected observed fluxes are used to evaluate
the intrinsic extinction. From these values, intrinsic extinction at $J$ is
determined assuming that it follows the Galactic extinction law given by Mathis
(1990). The extinction corrected flux at $J$ and the observed \mbox{X-ray} flux
are used to derive modified value of $\beta$. This procedure is iterated till
self consistent values of intrinsic extinction at $BVRI$ and $J$ wavelengths are
obtained. This yields $E(B-V) = 0.36\pm0.02$ mag and $A_J = 0.31\pm0.04$ as the 
values for the intrinsic extinction. These are in agreement with the values
determined independently by Price et al (2000b) from their optical data.
The observed fluxes corrected for both
Galactic and intrinsic extinction are also shown in Fig. 3. The values of
$\beta$ derived from them are $-0.94\pm0.02$ and $-1.02\pm0.02$ at $\Delta t =$
2.26 and 2.77 days respectively. 

In the light of above, we conclude that the value of $\beta$ is $\sim -0.95$ 
in  \mbox{X-ray} to near-IR spectral region and it has not changed before and
after $t_b$ during $\Delta t=$ 1 to 4 days.
 
Fig. 3 indicates that the peak frequency appears to lie in the millimeter 
region. This peak frequency is thus similar to that of GRB 970508 (cf. Galama 
et al. 1998) and GRB 000301C (cf. Sagar et al. 2000b) but different from that 
of GRB 990123 (Galama et al. 1999) where the peak is in radio region and that 
of GRB 971214 for which the peak is in optical/near-IR waveband (Ramaprakash 
et al. 1998). From this, one may infer that the synchrotron peak frequency may 
span a large range in GRB afterglows.

\begin{figure*}
\begin{center}\vspace*{-0.5cm}\hspace*{-1cm}
\epsfig{file=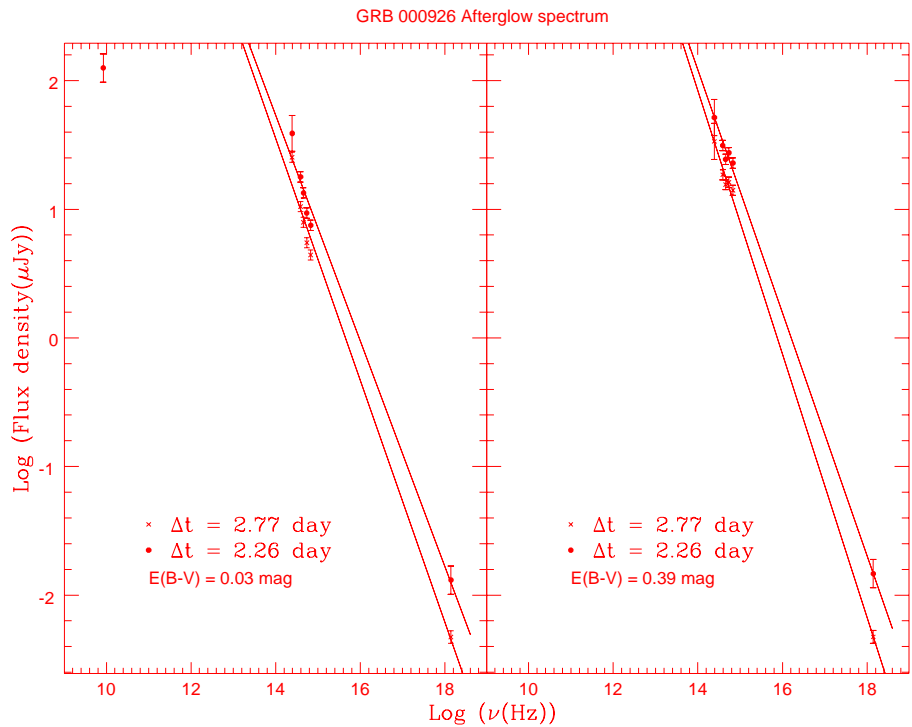,width=10.0cm,angle=270}
\end{center}\vspace*{-0.6cm}
\label{spec} 
\caption{The spectral flux distribution of the GRB 000926 
afterglow at $\sim$ 2.26 and 2.77 days after the burst.  The 
least square linear relations derived using fluxes at \mbox{X-ray}, $BVRI$ 
optical and $J$ near-IR are shown by solid lines for both epochs. Effect of
intrinsic extinction on the spectral slope can be seen from the two plots. } 
\end{figure*}

\section{Comparison with the Fireball model} 

The afterglow of a GRB is thought to be synchrotron radiation from a 
relativistic shock driven into the circumburst environment. In this model, the 
light curves and spectral energy distributions are generally well fit by the 
power-law  $F(\nu,t) \propto t^{-\alpha} \nu ^{\beta}$ for a range of 
frequencies and times that contain no spectral breaks.  The decay slope
is dependent on the dynamics of the fireball, and the spectral slope on the
importance of cooling, but within each regime $\alpha$ and $\beta$ are 
functions only of $p$, the power-law exponent of the electron Lorentz factor. 
This provides a verifiable relation between the two observable quantities 
$\alpha$ and $\beta$, and a way to infer the fireball dynamics. 

There is recent evidence that the fireball in at least some GRBs are not 
spherical but are collimated, like jets, into a small solid angle.
In such cases, theoretical models predict a break and a marked steepening 
in the afterglow light curve (M\'{e}sz\'{a}ros \& Rees 1999; Rhoads 1999; 
Sari et al. 1999, Huang 2000). The time of occurrence of this break in the 
afterglow light curve depends upon the opening angle of the collimated outflow. 
Observational evidence for a break was found first in the optical light curve 
of GRB 990123 afterglow (Castro-Tirado et al. 1999; Kulkarni et al. 1999) and
recently in that of GRB 990510 (Stanek et al. 1999), GRB 990705 (Masetti et al. 
2000), GRB 991216 (Halpern et al. 2000c) and GRB 000301C (Sagar et al. 2000b) 
afterglows. At late times, when the evolution is dominated by the lateral 
spreading of the jet, the value of 
$\alpha$ is expected to approach the electron energy distribution index  
while the value of $\beta$ is expected to be $-\alpha/2$ if the
cooling frequency is below the observing frequency and $-(\alpha-1)/2$ 
otherwise (Sari et al. 1999). The expected values of $\alpha$ and $\beta$ at 
late times are thus in agreement with the corresponding observed 
values of $2.6\pm0.1$ and $\sim -0.95$ for GRB 000926 indicating that the
afterglow emission from GRB 000926 is of jet type and not spherical, and that
the cooling frequency is above the range of observed frequencies.  The
slope of the light curve before the break is then expected to be $3\beta/2$
(Sari et al. 1998), which is also consistent with the value $1.45\pm0.05$ 
derived by us.  However, a much steeper late decay of \mbox{X-ray} flux with 
$\alpha = 4.3\pm$1 and $\sim$ 5, while the spectral index is the same 
($\sim -0.8$), as reported by Piro \& Antonelli (2000) and 
Garmire et al. (2000) respectively is not consistent with the simple jet model 
which predicts similar late flux decay in both optical and \mbox{X-ray} 
regions. It is possible that the geometry of the region emitting X-rays
after the lateral spreading starts differs from that emitting at longer
wavelengths, thus giving rise to apparent difference in the behaviour of
the light curve.  

Another area of possible disagreement with the standard
fireball model is the sharpness of the break in the light curve.  
While the expected sharpness of the transition could depend on the density
profile of the ambient medium (Kumar and Panaitescu 2000), the very
sharp transitions seen in GRB~000301C and GRB~000926 would be difficult to
quantitatively explain in the standard fireball model.  This is an area that
deserves a detailed theoretical study.

   The anisotropy of the initial ejection needs to be incorporated into
the derived energetics of the burst. The redshift $z = 2.0369$ determined
for the GRB~000926 afterglow by Castro et al. (2000) yields a minimum 
luminosity distance of 16.6 Gpc for standard Friedmann cosmology with
Hubble constant $H_0$ = 65 km/s/Mpc, cosmological density parameter 
$\Omega_0$ = 0.2 and cosmological constant $\Lambda_0$ = 0 (if $\Lambda_0 > 0$ 
then the inferred distance would increase).
The GRB~000926 thus becomes the second farthest GRB after GRB~971214 (Kulkarni 
et al. 1998) amongst the GRBs with known redshift measurements so far. 
If the original emission were isotropic, the observed fluence of $2.2 \times 
10^{-5}$~erg/cm$^2$ (Hurley et al. 2000) between 25--100 keV, would yield the
total $\gamma-$ray energy release to be at least $2.5 \times 10^{53}$ erg 
($\sim 0.14 M_{\odot}c^2$). However, the estimated $t_b$ of 1.74 day implies,
using the expression in Sari, Piran and Halpern (1999), a jet opening angle 
of $0.14 n^{1/8}$ radian, where $n$ is the number density of the ambient medium.
This means that the actual energy 
released from the GRB 000926 is reduced by a factor of $\sim 100$ relative to 
the isotropic value and becomes $\sim 2 \times 10^{51}$ ergs.

Of the over dozen GRBs with known redshifts, seven with inferred energy budget
of $> 10^{53}$ erg (assuming isotropic emission) are GRB~000926 (discussed 
here); GRB 000301C  (Sagar et al. 2000b); GRB 991216 and GRB 
991208 (Sagar et al. 2000a); GRB 990510 (Harrison et al. 1999); GRB 990123 
(Andersen et al. 1999; Galama et al. 1999) and  GRB 971214 (Kulkarni et al. 
1998). Recent observations suggest that GRBs are associated with stellar deaths,
and not with quasars or the nuclei of galaxies as some GRBs are found offset 
by a median value of 3.1 kpc from the centre of their host galaxy (cf. Bloom et 
al. 2000). However, release of energies as large as $\sim 10^{53}$ erg or more 
in radiation is extremely difficult to accommodate within the 
popular stellar death models (coalescence of neutron stars or the death of 
massive stars). However, evidence is now mounting in favour of non-isotropic
emission, with GRB~000926 the sixth known case after GRB 990123 
(cf.\ Castro-Tirado et al.\ 1999; Kulkarni et al.\ 1999), GRB 990510 
(cf.\ Harrison et al.\ 1999, Stanek et al.\ 1999), GRB 990705 
(cf.\ Masetti et al.\ 2000), GRB 991216 (cf.\ Halpern et al.\ 2000c, 
Sagar et al.\ 2000a) and GRB 000301C (cf.\ Sagar et al.\ 2000b).

Thus, it is possible that all cases where the inferred isotropic equivalent
energy is large, the emission actually is confined within narrow jets, with
total energy much below the isotropic equivalent.  The $\gamma-$ray energy 
released then becomes $\leq 10^{52}$ erg, a value within the reach of the 
currently popular models for the origin of GRBs (see Piran 1999 and references 
therein).

The intrinsic visual extinction ($A_V \sim 1.1$ mag) derived for the 
optical afterglow also carries a possible clue to the class of progenitor
objects.  Extinction of comparable magnitude have been noticed also in 
GRB 971214 ($A_V \sim 0.9$ mag, Ramaprakash et al.\ 1998 and refernces therein);
GRB 980329 ($A_V \sim 1.2$ mag, Palazzi et al.\ 1998);
GRB 980703 ($A_V \sim 0.8 - 1.2$ mag, Vreeswijk et al.\ 1999) and
GRB 000418 ($A_V \sim 1.0$ mag, Klose et al.\ 2000).
The presence of dust extinction in the host galaxies broadly supports the
proposal that GRBs could be associated with massive stars embedded in
star-forming regions of the GRB host galaxies (Paczy\'{n}ski 1998).

\section{Conclusions }

We have determined $BVRI$ magnitudes of 20 stars in the field of GRB 000926. 
These magnitudes are used to calibrate present photometric magnitudes of the OT 
of GRB 000926 as well as those given in GCN circulars using differential 
photometric techniques.  The light curves show a steepening which could be 
well understood in terms of a jet model. The parameters of the jet are derived 
by performing a non-linear least square fit of a broken power-law model to the 
light curves. The flux decay constants at early and late times are 
1.45$\pm$0.05 and 2.6$\pm$0.1 respectively, and the break occurs at 
1.74$\pm$0.11 day.  The break is sharp and similar to that in GRB 000301C 
(cf. Sagar et al. 2000b). 

The quasi-simultaneous spectral energy distributions 
determined for various epochs indicate that the spectral index of the 
GRB 000926 afterglow has not changed significantly during this period. A
steepening of the flux decay with no corresponding change in spectral index is 
expected of a jet evolution.

The determined redshift yields a minimum distance of 16.6 Gpc to GRB~000926,
and the observed fluence yields an isotropic equivalent energy release of 
$\sim 2.5\times 10^{53}$~erg.  However, the break in the light curve at 1.74
days translate to a collimated emission with an opening angle of $\sim 0.14$
radian, reducing the estimated energy release to $\sim 2\times 10^{51}$ erg.

The late time flux decays at \mbox{X-ray} and optical wavelengths seem to be 
significantly different for GRB 000926 afterglow. Similarly, the flux decays 
of GRB 991216 afterglow (cf. Frail et al. 2000) are also different at 
\mbox{X-ray}, optical and radio wavelengths. The multi-wavelength observations 
of recent GRB afterglows have thus started revealing features which require 
explanations other than generally accepted so far indicating that there may 
be yet new surprises in GRB afterglows.

\bigskip
\noindent {\bf Acknowledgements:} This research has made use of data 
obtained through the High Energy Astrophysics Science Archive Research Center 
Online Service, provided by the NASA/Goddard Space Flight Center.

\medskip
\noindent {\bf References:} 
\begin{itemize}
\item [] Andersen M. I.  et al., 1999, Science, {\bf 283}, 2075
\item [] Bessell M.S., Brett J.M., 1988, PASP, {\bf 100}, 1134
\item [] Bloom J.S., Kulkarni S.R., Djorgovski S.G., 2000, astro-ph/0010176
\item [] Castro S.M. et al., 2000, GCN Observational Report No. 851
\item [] Castro-Tirado A. J.  et al., 1999, Science, {\bf 283}, 2069
\item [] Dall T. et al., 2000, GCN Observational Report No. 804
\item [] Di Paola A. et al., 2000, GCN Observational Report No. 816
\item [] Frail D.A., Berger E., 2000, GCN Observational Report No. 805
\item [] Frail D.A. et al., 2000, ApJ, {\bf 538}, L129/astro-ph/0003138
\item [] Fynbo  J.P.U. et al., 2000a, GCN Observational Report No. 820
\item [] Fynbo  J.P.U. et al., 2000b, GCN Observational Report No. 825
\item [] Fynbo  J.P.U. et al., 2000c, GCN Observational Report No. 807
\item [] Fynbo  J.P.U. et al., 2000d, GCN Observational Report No. 840
\item [] Fynbo  J.P.U. et al., 2000e, GCN Observational Report No. 871
\item [] Fukugita  M., Shimasaku K., Ichikawa T., 1995, PASP, {\bf 107}, 945
\item [] Galama  T.J., 2000, to appear in Proc. of the 5th Huntsville Gamma-Ray 
         Burst Symposium/astro-ph/0001465
\item [] Galama  T.J. et al., 1998, ApJ, {\bf 500}, L97
\item [] Galama T.J. et al., 1999, Nature, {\bf 398}, 394
\item [] Garmire G., Garmire A., Piro L., Garcia M.R.,  2000, GCN Observational 
         Report No. 836
\item [] Gorosabel J. et al., 2000, GCN Observational Report No. 803
\item [] Halpern J.P. et al., 2000a, GCN Observational Report No. 824
\item [] Halpern J.P. et al., 2000b, GCN Observational Report No. 829
\item [] Halpern J.P. et al., 2000c, ApJ, {\bf 543}, 697
\item [] Harisson F.A. et al., 1999, ApJ, {\bf 523}, L121
\item [] Hjorth J. et al., 2000a, GCN Observational Report No. 809
\item [] Hjorth J. et al., 2000b, GCN Observational Report No. 814
\item [] Huang Y.F., Dai Z.G., Lu T., 2000, MNRAS, 316, 943
\item [] Hurley K., Mazets E., Golenetskii S., 2000, GCN Observational Report 
         No. 801, 802
\item [] Klose S., 2000, Rev. Modern Phys. {\bf 13}, 129
\item [] Klose S. et al, 2000, ApJ, {\bf 545}, 271
\item [] Kobayashi N. et al., 2000, GCN Observational Report No. 821
\item [] Kulkarni S.R. et al., 1998, Nature, {\bf 393}, 35
\item [] Kulkarni S.R. et al., 1999, Nature, {\bf 398}, 389
\item [] Kulkarni S.R. et al., 2000, to appear in Proc. of the 5th Huntsville 
         Gamma-Ray Burst Symposium/astro-ph/0002168
\item [] Kumar P., Panaitescu A., 2000, ApJ, {\bf 541}, L9
\item [] Lamb, D.Q., 2000, astro-ph/0005028
\item [] Landolt, A.R., 1992, AJ, {\bf 104}, 340
\item [] Masetti N. et al., 2000, A\&A, 354, 473
\item [] Mathis J.S., 1990, ARAA, {\bf 28,} 37
\item [] M\'{e}sz\'{a}ros P., Rees M. J., 1999, MNRAS, {\bf 306}, L39
\item [] Monet D., 1997, The PMM USNO-A1.0 catalogue
\item [] Paczy\'{n}ski B., 1998, ApJ, {\bf 494}, L45
\item [] Palazzi E. et al., 1998, A\&A, {\bf 336}, L95
\item [] Piran T., 1999, Physics Reports {\bf 314},  575 
\item [] Piro  L., Antonelli L.A., 2000, GCN Observational Report No. 832, 833
\item [] Price  P.A. et al., 2000a, GCN Observational Report No. 811
\item [] Price  P.A. et al., 2000b, Submitted to ApJ Letters,/astro-ph/0012303
\item [] Ramaprakash, A.N. et al., 1998, Nature, {\bf 393}, 43
\item [] Rhoads J.E., 1999, ApJ, {\bf 525}, 737
\item [] Rol E., Vreeswijk P.M., Tanvir N., 2000, GCN Observational Report 
         No. 850
\item [] Sagar R., Pandey A.K, Mohan V., Yadav R.K.S., Nilakshi, 
         Bhattacharya D., Castro-Tirado A.J.,  1999, BASI, {\bf 27}, 
         3/astro-ph/9902196
\item [] Sagar R., Mohan V., Pandey A.K., Pandey S.B., Castro-Tirado A.J., 
         2000a, BASI, {\bf 28}, 15/astro-ph/0003257
\item [] Sagar R., Mohan V., Pandey S.B., Pandey A.K., Stalin C.S.,
         Castro-Tirado A.J., 2000b, BASI, {\bf 28}, 499/astro-ph/0004223
\item [] Sari R., Piran T., Halpern J. P., 1999, ApJ, {\bf 519}, L17
\item [] Sari R., Piran T., Narayan R., 1998, ApJ, {\bf 497}, L17
\item [] Schlegel D.J., Finkbeiner D.P., Davis M., 1998, ApJ, {\bf 500}, 525
\item [] Stanek K. Z. et al., 1999, ApJ {\bf 522}, L39
\item [] Veillet C., 2000, GCN Observational Report No. 831
\item [] Vrba F., Canzian B., 2000, GCN Observational Report No. 819
\item [] Vreeswijk P.M. et al., 1999, ApJ, {\bf 523}, 171
\end{itemize}
\end{document}